\documentclass[%
  reprint,
  amsmath,amssymb,
  aps, prl,
  floatfix,
]{revtex4-2}
\usepackage{color}
\usepackage{xcolor}
\usepackage{graphicx}
\usepackage{dcolumn}
\usepackage{bm}
\usepackage{cancel}

\begin{document}

\preprint{APS/123-QED}

\title{Spatiotemporally correlated activity regulation of random-organizing hyperuniform states}

\author{Yuhan Peng}
\thanks{These authors contributed equally.}

\author{Yusheng Lei}
\thanks{These authors contributed equally.}

\author{Ran Ni}
\email{r.ni@ntu.edu.sg}

\affiliation{School of Chemistry, Chemical Engineering and Biotechnology, Nanyang Technological University, 62 Nanyang Drive, Singapore 637459}

\date{\today}

\begin{abstract}
  Center-of-mass (COM) conservation has been widely accepted as an important mechanism for generating nonequilibrium dynamic hyperuniform states with vanishing long-wavelength density fluctuations, but whether it is sufficient for hyperuniformity remains unclear. Here, using particle simulations and a linear fluctuating theory, we show that spatiotemporally correlated activity regulation can destroy hyperuniformity while preserving COM conservation. We construct a binary mixture in which the local density of independently evolving driver particles regulates the activity of regulated particles following COM-conserving biased random organization. This counterexample demonstrates that hyperuniformity depends on the spatial and temporal correlations of activity as well as the conservation laws governing particle displacements. Our findings clarify the limits of COM conservation as a route to hyperuniformity and highlight the importance of controlling activity correlations in the design of nonequilibrium dynamic hyperuniform mixtures.
\end{abstract}


\maketitle

\paragraph{Introduction.}
A statistically homogeneous particle system is hyperuniform if the variance of the number of particles in a large sampling window grows more slowly than the window volume, corresponding to a vanishing structure factor in the long-wavelength limit, $S(q\to0)=0$ with $q=|\boldsymbol q|$ the wavevector magnitude~\cite{torquato2003local,torquato2018hyperuniform}. Examples of hyperuniform structures include photoreceptor cell patterns in birds~\cite{jiao2014avian}\nocite{kram2010avian,yellott1983spectral}, eigenvalues of random matrices~\cite{ginibre1965statistical}, maximally random jammed packings~\cite{donev2005unexpected}, low-discrepancy sequences~\cite{kuipers2012uniform}, and early universe fluctuations~\cite{gabrielli2002glass}, etc. These findings motivate the search for physical processes that generate hyperuniformity, which can be used to design novel disordered photonic materials~\cite{florescu2009designer,man2013isotropic,leseur2016high}.
In equilibrium, hyperuniformity can be achieved through long-range interactions, like in Coulomb plasmas~\cite{torquato2003local,diamant2025fluids,wang2024designer,Zhang2020Realizable}. However, such long-range interactions are not necessary out of equilibrium, where short-range particle interactions can produce hyperuniformity through the dynamics itself~\cite{torquato2018hyperuniform,lei2025non}. This opens promising routes to hyperuniform states via nonequilibrium dynamics.
Absorbing transitions produce critical hyperuniformity in random organization (RO)~\cite{pine2005chaos,corte2008random,hexner2015hyperuniformity,tjhung2015hyperuniform,ma2019generalized,wilken2020hyperuniform} and nonreciprocal active mixtures~\cite{chen2024emergent}, with related metastable hyperuniformity near discontinuous transitions~\cite{lei2023howHUfreeze,lei2026metastable}.
Away from criticality, hyperuniformity arises in RO models with center-of-mass (COM) conservation~\cite{hexner2017noise,anand2026universal}, random-organizing hard-sphere fluids~\cite{lei2019hydrodynamics}, pulsating cell tissues~\cite{li2025fluidization}, and chiral active particles~\cite{lei2019nonequilibrium,kuroda2023microscopic}, which was observed experimentally in chiral Quincke rollers~\cite{zhangHyperuniformActiveChiral2022}.




Previous studies have established a close connection between COM conservation and hyperuniformity. In modified RO, Manna models, and their generalizations, COM conservation provides a physical explanation for active-state hyperuniformity~\cite{hexner2017noise,lei2019nonequilibrium,kuroda2023microscopic,delucaHyperuniformityPhaseOrdering2024a,li2025fluidization,maire2025hyperuniformity,anand2026universal,casiulis2026hyperuniform,lei2023sphere,maire2024enhancing,galliano2023crystals,maire2025interfaces,maire2025granular,kuroda2025crystal,gao2026criticality,maire2026nucleation,bertrand2019nonlinear,tjhung2017discontinuous}. Recent work further points out the joint role of conservation laws and nonequilibrium driving to hyperuniformity~\cite{maire2025hyperuniformity}. However, whether COM conservation is sufficient to ensure hyperuniformity out of equilibrium still remains unclear.
In this Letter, we show that COM conservation alone does not guarantee hyperuniformity. To show this, we construct a counterexample consisting of an $AB$ binary mixture. Species $B$ follows COM-conserving biased random organization (BRO), while its activity is regulated by an independently evolving species $A$. Computer simulations show a nonzero small-$q$ plateau in the structure factor of COM-conserving species $B$, indicating the loss of hyperuniformity. To understand this, we develop a linear fluctuating theory to explain how the spatiotemporal correlation of the species $A$ density fluctuation regulates large-scale density fluctuations in species $B$.

\begin{figure}[!t]
  \centering
  \includegraphics[width=\columnwidth]{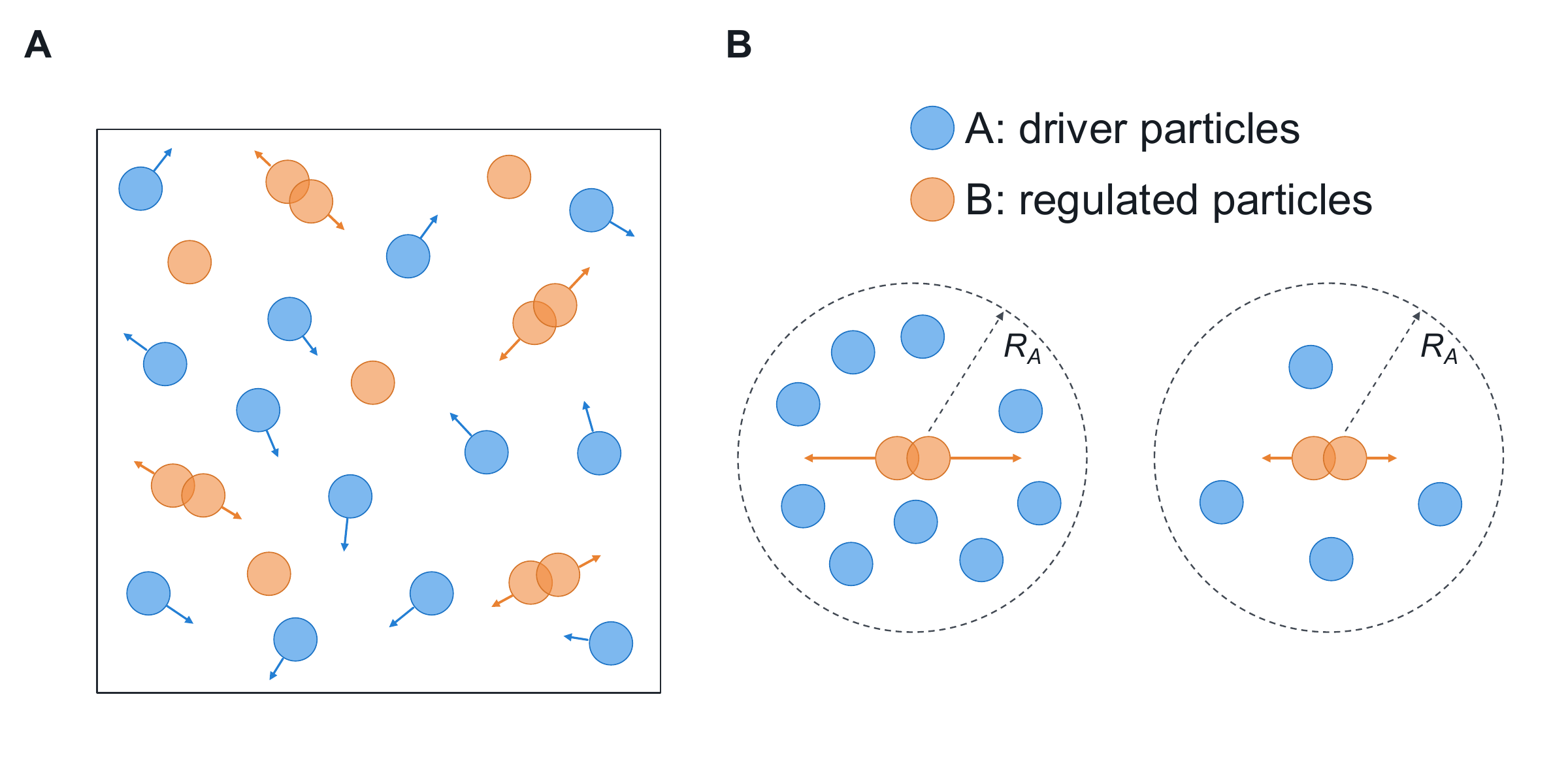}
  \caption{\textbf{Activity regulated COM-conserving biased random organization.}
    \textbf{(A)} Driver particles $A$ (blue) evolve independently,
    while regulated BRO particles $B$ (orange) receive active kicks only when
    overlapping with another $B$ particle. Darker shading indicates overlaps. Note that the activity regulation is nonreciprocal: species $A$ affects the activity of species $B$, while $B$ has no effect on $A$.
    \textbf{(B)} The local driver density is evaluated within a circular
    sensing region of radius $R_A$ centered at the midpoint of each
    overlapping $B$--$B$ pair. Dashed circles delineate sensing regions,
    not physical boundaries. The pair receives equal and opposite
    displacements along its line of centers, each with magnitude regulated by the local $A$ particle density fluctuation.
    Here higher local driver density increases activity while preserving pairwise COM conservation.}
  \label{fig:schematic}
\end{figure}

\paragraph{Model.}
We consider a two-dimensional mixture of species $A$ and $B$, with
equal particle diameters $\sigma$, in a periodic square simulation box with system size $L$. The particle numbers $N_A$ and $N_B$ are fixed, giving mean densities $\bar\rho_A=N_A/L^2$ and $\bar\rho_B=N_B/L^2$.
In the system, species $A$, driver particles, evolves independently, while its local density regulates the activity of species $B$, regulated particles (Fig.~\ref{fig:schematic}). Regulated $B$ particles follow COM-conserving BRO with reciprocal displacements as defined in Ref.~\cite{casiulis2026hyperuniform}. We distinguish RO, with random displacement directions and magnitudes, from BRO, with random displacement magnitudes along the line of centers of overlapping particles. In this work, regulated $B$ particles are hyperuniform in the active state with $S(q \to 0)\sim q^2$ in the absence of the regulation by driver particles $A$~\cite{hexner2017noise}.
Specifically, at each simulation step, all overlapping $B$--$B$ pairs receive active kicks according to the protocol below, while the other $B$ particles remain immobile.
For each overlapping species $B$ pair $(i,j)$, we count the number $n_A^{(ij)}$ of $A$ particles within a circular sensing window of radius $R_A$ centered at the midpoint of the pair $ij$. The local density deviation is $\delta\rho_A^{(ij)}=n_A^{(ij)}/(\pi R_A^2)-\bar\rho_A$, and the species $B$ pair receives equal and opposite displacements regulated by local density of species $A$,
\begin{equation}
  \Delta\mathbf{r}_i^{(ij)}=-u_{ij}\sigma(h_0+h\delta\rho_A^{(ij)})\hat{\mathbf{e}}_{ij},
  \qquad
  \Delta\mathbf{r}_j^{(ij)}=-\Delta\mathbf{r}_i^{(ij)},
  \label{eq:pair_kicks}
\end{equation}
with $\hat{\mathbf e}_{ij}$ the unit vector from $i$ to $j$ and
$u_{ij}$ an independent random number drawn uniformly from $[0,1)$. Here $h_0$ sets the intrinsic activity and $h$ controls the additional activity from regulation.
In this work, systems with an ideal-gas driver or a BRO driver are investigated. We use $L=300\sigma$, $R_A=3\sigma$, and initialize both species at independent and random positions. The mean number densities are $\bar\rho_A=\bar\rho_B=0.6\sigma^{-2}$ for the ideal-gas driver and $\bar\rho_A=0.8\sigma^{-2}$, $\bar\rho_B=0.7\sigma^{-2}$ for the BRO driver.
Each simulation runs for $10^5$ steps, with configurations sampled during the second half of the run (SI).

We study the impact of the regulation by driver particles $A$ on the structure of regulated $B$ particles. Here, two types of drivers are used to distinguish the roles of temporal persistence and spatial correlations of the activity of $B$ particles, regulated by driver particles $A$, in the loss of hyperuniformity in $B$ particles. The ideal-gas driver isolates temporal persistence changes from fixed spatial correlation, whereas the BRO driver primarily probes spatial correlation changes at a nearly fixed temporal persistence. In both cases, the displacements of $B$ remain reciprocal, allowing us to examine how the correlations of their activity affect density fluctuations under the same COM conservation law.


\begin{figure*}[!t]
  \centering
  \makebox[\textwidth][c]{\includegraphics[width=0.95\textwidth]{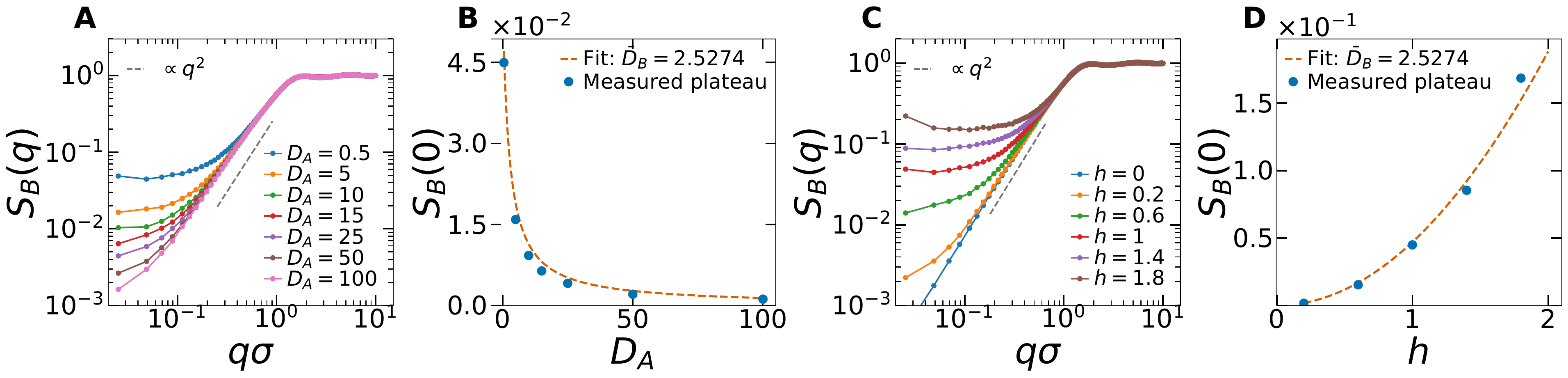}}
  \caption{\textbf{Density fluctuations of regulated particles under an ideal-gas driver.}
    \textbf{(A)} Structure factor $S_B(q)$ for driver diffusivities
    $D_A$ from $0.5$ to $100$ at regulation strength $h=1$.
    \textbf{(B)} Estimated long-wavelength plateau $S_B(0)$ as a function of $D_A$ at $h=1$. See SI for plateau measurement details.
    \textbf{(C)} $S_B(q)$ for $h$ from $0$ to $1.8$ at $D_A=0.5$.
    \textbf{(D)} Estimated plateau $S_B(0)$ as a function of $h$
    at $D_A=0.5$.
    Symbols show simulation data. The intrinsic strength $h_0$ for $B$ is $2$. Orange dashed curves in (B,D) show the fit to Eq.~\eqref{eq:transfer_spectrum}; gray dashed
    lines in (A,C) are guides to $q^2$ scaling.
  }
  \label{fig:ig}
\end{figure*}
\paragraph{Ideal-gas driver.}
In this case, driver particles $A$ undergo Brownian dynamics without interactions, so $S_A(q)=1$ for all driver diffusivities $D_A$ (see SI). Increasing $D_A$ shortens the temporal correlations of the driver density without changing its spatial correlations. This allows us to evaluate the effect of the temporal persistence of activity, regulated by drivers $A$, on density fluctuations in $B$.
At fixed nonzero regulation strength $h$, $S_B(q)$ follows $q^2$ scaling at intermediate wavevectors but approaches a finite plateau at small $q$ (Fig.~\ref{fig:ig}(A)).
The $q^2$ regime indicates a hyperuniform fluctuation suppression at the intermediate length scales, and the finite plateau at larger length-scale, indicates a loss of hyperuniformity despite the fluctuation suppression at the intermediate wavelengths and the COM conservation.
Increasing $D_A$ at fixed $h$ lowers the plateau (Fig.~\ref{fig:ig}(B)). A lower plateau means weaker large-scale density fluctuations, i.e., a more uniform structure of $B$ on large length-scales. In the limit of $D_A\to\infty$, which corresponds to no temporal persistence, the system is expected to be hyperuniform.
At fixed $D_A$, increasing $h$ raises the small-$q$ plateau (Fig.~\ref{fig:ig}(C)). The plateau grows
approximately quadratically with $h$ (Fig.~\ref{fig:ig}(D)), whereas
the large-$q$ structures remain nearly intact. Stronger regulation
therefore corresponds to more significant long-wavelength density fluctuations,
with little change on small length-scales.  The orange dashed curve is a fit to the theory developed later.

\begin{figure*}[!t]
  \centering
  \makebox[\textwidth][c]{\includegraphics[width=0.95\textwidth]{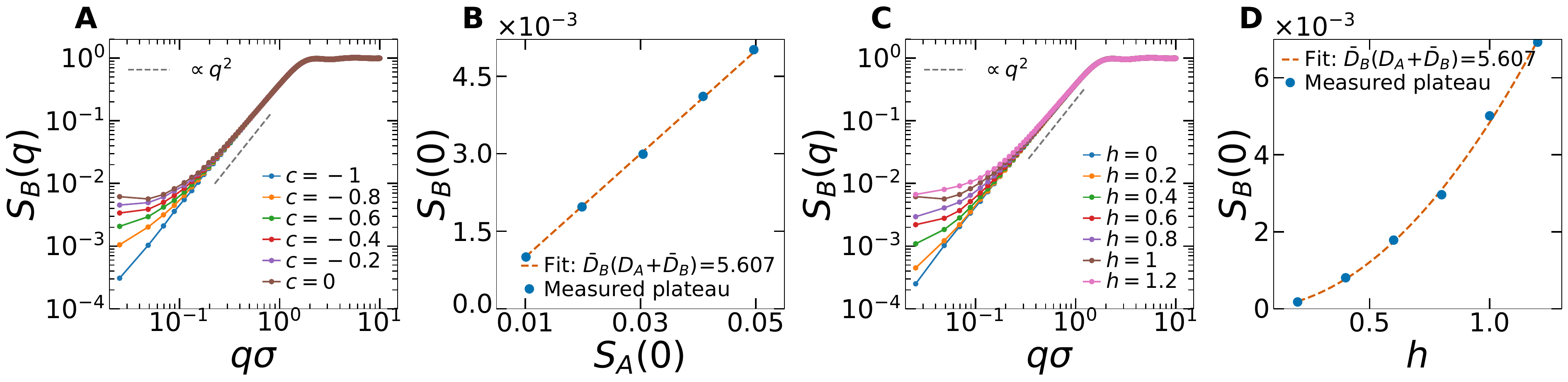}}
  \caption{\textbf{Density fluctuations of regulated particles under a BRO driver.}
    \textbf{(A)} Structure factor $S_B(q)$ for different values of
    the driver correlation parameter $c$ at regulation strength $h=1$.
    \textbf{(B)} Estimated long-wavelength plateau $S_B(0)$ as a
    function of the measured driver plateau $S_A(0)$ at $h=1$.
    \textbf{(C)} $S_B(q)$ for $h$ from $0$ to $1.2$ at $c=0$.
    \textbf{(D)} Estimated plateau $S_B(0)$ as a function of $h$
    at $c=0$.
    Symbols are from simulation. The intrinsic strengths $h_0$ for $B$ is $1.5$, while the activity strength for $A$ is $1$. Orange dashed curves in (B,D) show the
    fit to Eq.~\eqref{eq:transfer_spectrum}; gray dashed
    lines in (A,C) are guides to $q^2$ scaling.
  }
  \label{fig:bro}
\end{figure*}

\paragraph{Biased random organization driver.}
To examine how spatial correlations in the driver affect the structure of regulated $B$ particles, we model species $A$ using BRO~\cite{casiulis2026hyperuniform}. Its nonequilibrium dynamics enables us to tune the structure independently of the temporal persistence, free from the constraint of the Einstein relation.
Overlapped $A$ particles receive active kicks along their line of centers, with uniformly distributed magnitudes. The parameter $c\in[-1,0]$ controls how the two kick magnitudes are sampled. With probability $-c$, both particles receive the same magnitude; otherwise, their magnitudes are sampled independently. Thus, $c=-1$ conserves the COM of each pair, while $c=0$ gives independent magnitudes. The choice of $c$ has a great impact on $S_A(q)$. Increasing $c$ from $-1$ to $0$ raises the plateau $S_A(q \to 0)$ from zero to progressively larger values, while the measured particle diffusivity of $A$ changes only weakly for all $c$ (see SI).
When driver kick magnitudes are strictly reciprocal ($c=-1$), pairwise center-of-mass conservation in species $A$ suppresses its long-wavelength density fluctuations, yielding an active hyperuniform driver with $S_A(q \to 0) \sim q^2$ (see SI). Under this driver, species $B$ preserves its intrinsic hyperuniformity, scaling as $S_B(q) \propto q^2$ as $q \to 0$ (Fig.~\ref{fig:bro}(A)).
As $c$ goes to $ 0$, a low-$q$ plateau emerges and grows in $S_A(q \to 0)$, which subsequently induces a corresponding plateau in species $B$ at small wavevectors. This behavior explicitly shows how spatial correlations of activity, regulated by drivers $A$, affect the structure of regulated $B$ particles. Quantitatively, $S_B(q \to 0)$ grows approximately linearly with the measured $S_A(0)$ (Fig.~\ref{fig:bro}(B)).

To clarify the dependence of species $B$'s structure on the coupling strength, we fix $c=0$ and vary $h$ (Fig.~\ref{fig:bro}(C,D)). At $c=0$, increasing $h$ raises the small-$q$ plateau approximately quadratically while leaving the large-$q$ structure nearly unaffected, like the ideal-gas driver (Fig.~\ref{fig:bro}(C,D)). The crossover to $q^2$ scaling shifts to larger $q$, narrowing the range of fluctuation suppression.
Together, these results show that the long-wavelength structure of $B$ varies with the driver dynamics, the driver structure, and the regulation strength, while all $B$--$B$ pair displacements remain COM-conserving. The observed low-$q$ plateaus demonstrate that COM conservation alone does not ensure hyperuniformity. We next develop a linear fluctuating theory to account for these trends.

\textit{Theoretical analysis.---}
To understand the large-scale structure of $B$ under nonreciprocal activity regulation by A, we formulate a linear fluctuating theory around the homogeneous active steady state. The equations for the density fluctuations can be written as (see SI)
\begin{equation}
  \begin{aligned}
    \partial_t\delta\rho_A={}&D_A\nabla^2\delta\rho_A
    +A_1\sqrt{\bar\rho_A}\nabla^2\eta_1
    +A_2\sqrt{\bar\rho_A}\nabla\cdot\boldsymbol\eta_2,\\
    \partial_t\delta\rho_B={}&\bar D_B\nabla^2\delta\rho_B
    +h\bar\rho_B\nabla^2\delta\rho_A
    +\bar B\sqrt{\bar\rho_B}\nabla^2\eta_3,
  \end{aligned}
  \label{eq:theory_B}
\end{equation}
where $\delta\rho_{\alpha}=\rho_{\alpha}-\bar\rho_{\alpha}$ ($\alpha=A,B$) is the local
number-density fluctuation with $\bar\rho_{\alpha}$ the mean density. The first two terms on the right hand side of the species $B$ equation come from the activity-driven diffusion $\nabla^2\left[D_B(\rho_A )\rho_B\right]$, which preserves COM conservation. 
To describe the activity regulation of $B$ by $A$, we approximate local diffusivity of $B$ as a linear function of $A$'s density fluctuation: $D_B(\rho_A)=\bar D_B+h\delta\rho_A$, where $h$ measures the regulation strength. The diffusivity constants $D_A$ and $\bar D_B$ describe the relaxation of density fluctuations of the two species. 
$\eta_1$, $\boldsymbol\eta_2$ and $\eta_3$ are independent Gaussian white noises with zero
mean and unit strength; $A_1$, $A_2$, and $\bar B$ set their noise amplitudes. The Laplacian noise terms $\nabla^2\eta_1$ and $\nabla^2\eta_3$ conserve particle number and COM, whereas the diffusive noise term $\nabla\cdot\boldsymbol\eta_2$
conserves particle number only.

By solving Eq.~\eqref{eq:theory_B} in Fourier space with $q=|\boldsymbol q|$, we obtain the static structure factors
\begin{equation}
  \begin{aligned}
    S_A(q)&=\frac{A_2^2+A_1^2q^2}{2D_A},\\
    S_B(q)&=\frac{\bar B^2}{2\bar D_B}q^2
    +\frac{h^2\bar\rho_A\bar\rho_B}
    {\bar D_B(D_A+\bar D_B)}S_A(q).
  \end{aligned}
  \label{eq:transfer_spectrum}
\end{equation}
The first term in $S_B(q)$ describes the intrinsic density fluctuations
of $B$, while the second term is induced by activity regulation and is proportional
to the driver's equal-time structure $S_A(q)$. For nonzero regulation strength, the induced contribution
therefore has the same small-$q$ dependence as $S_A(q)$ as shown in Eq.~\ref{eq:transfer_spectrum}.
For the ideal-gas driver, we have
$A_1=0$ and $A_2=\sqrt{2D_A}$, so $S_A(q)=1$ and
$S_B(q\to0)=h^2\bar\rho_A\bar\rho_B/
[\bar D_B(D_A+\bar D_B)]$. Increasing $D_A$ therefore lowers the plateau while the driver $A$'s structure remains intact, as shown in
Fig.~\ref{fig:ig}(B). For the BRO driver with $A_2=0$, both $S_A(q)$ and $S_B(q)$ scale as $q^2$, consistent with the $c=-1$ result. When $A_2>0$, regulation instead produces a plateau proportional to $h^2S_A(0)$ at fixed densities and diffusivities, where $S_A(0)$ denotes $S_A (q \to 0)$. This accounts for the linear dependence on the driver plateau and the quadratic dependence on $h$ in Fig.~\ref{fig:bro}(B,D).

To further understand the physical origin of the induced fluctuations, we consider how the density fluctuations of $A$ regulate the local activity of $B$. The resulting activity regulation inherits both the spatial correlations and the temporal persistence of the driver fluctuations. Through the activity regulation term, it acts on the density dynamics of $B$ as an effective noise $\Xi=h\sqrt{\bar\rho_B}\nabla^2\delta\rho_A$, which enters Eq. (\ref{eq:theory_B}) for $B$ as $\sqrt{\bar\rho_B}\Xi$.
The spatiotemporal correlations of the effective noise follow from the density dynamics of $A$ and are described by the intensity $\langle\tilde\Xi
\tilde\Xi^*\rangle\propto h^2\bar\rho_A\bar\rho_B q^4S_A(q,\omega)$, with the dynamic structure factor of $A$: $S_A(q,\omega)=D_Aq^2S_A(q)/[\pi(\omega^2+D_A^2q^4)]$ and $\tilde\Xi$ the Fourier transform of $\Xi$.
Here $S_A(q)$ characterizes the equal-time spatial correlations, while $D_A$ sets the correlation time $1/(D_Aq^2)$.
The Fourier and related conventions are given in the SI.

To separate the noise intensity from its temporal dependence, we introduce a white Gaussian reference noise $\Xi^{\mathrm{white}}$ whose power spectrum equals the zero-frequency limit of that of $\Xi$.
The covariance of $\Xi$ can be obtained as (see SI)
\begin{equation}
  \begin{aligned}
    &\langle\tilde\Xi(\boldsymbol q,\omega)
    \tilde\Xi^*(\boldsymbol q',\omega')\rangle\\
    &\quad=\frac{D_A^2q^4}{\omega^2+D_A^2q^4}
    \langle\tilde\Xi^{\mathrm{white}}(\boldsymbol q,\omega)
    \tilde\Xi^{\mathrm{white}*}(\boldsymbol q',\omega')\rangle.
  \end{aligned}
  \label{eq:theory_colored_noise}
\end{equation}
This representation retains the temporal correlations of $\Xi$ and does not require an adiabatic approximation. The intensity of the white reference noise is $\langle\tilde\Xi^{\mathrm{white}}(\boldsymbol q,\omega)
\tilde\Xi^{\mathrm{white}*}(\boldsymbol q',\omega')\rangle
=(2\pi)^3
\delta(\boldsymbol q-\boldsymbol q')
\delta(\omega-\omega')[2\pi h^2\bar\rho_A\bar\rho_Bq^4S_A(q,\omega\rightarrow 0)]$ with $A$'s dynamic structure factor $S_A(q,\omega\rightarrow 0)= S_A(q)/(\pi D_Aq^2)$, which measures the time-integrated behavior of the driver density fluctuations. It therefore depends on both the static correlation and temporal persistence of the fluctuations of $A$ that regulate the activity of $B$.

Combining the effective noise with  Eq.~\eqref{eq:theory_B} for $B$, we obtain
\begin{equation}
  \begin{aligned}
    &S_B(q,\omega)=\lim_{\tau_{\max}\to\infty}\frac{1}{2\pi V\tau_{\max}(\omega^2+\bar D_B^2q^4)}\\
    &\quad\times\Bigg[\bar B^2q^4\langle\tilde\eta_3\tilde\eta_3^*\rangle
      +\frac{D_A^2q^4}{\omega^2+D_A^2q^4}
    \langle\tilde\Xi^{\mathrm{white}}\tilde\Xi^{\mathrm{white}*}\rangle\Bigg].
  \end{aligned}
  \label{eq:theory_dynamic_structure}
\end{equation}
The intrinsic and effective noise contributions are both weighted
by the relaxation of density fluctuations in $B$.
Integrating Eq.~\eqref{eq:theory_dynamic_structure} over frequency gives
\begin{equation}
  S_B(q)=S_B^0(q)+\frac{D_A}{D_A+\bar D_B}S_B^{\mathrm{white}}(q),
  \label{eq:theory_noise_balance}
\end{equation}
where $S_B^0(q)=\bar B^2q^2/(2\bar D_B)$ is the intrinsic contribution and $S_B^{\mathrm{white}}(q)=h^2\bar\rho_A\bar\rho_BS_A(q)/(D_A\bar D_B)$ is the contribution generated by the white reference noise alone.


The contribution induced by activity regulation in Eq.~\eqref{eq:theory_noise_balance} is the product of the white reference $S_B^{\mathrm{white}}(q)$ and the temporal correction factor $D_A/(D_A+\bar D_B)$. Here, the white reference yields the structure under equivalent white noise $\Xi^{\mathrm{white}}$, while the correction factor captures the finite persistence of $\Xi$ relative to the relaxation of $B$. In the adiabatic limit, $D_A\gg D_B$, this factor approaches unity. Crucially, because it is independent of $q$, the white reference captures the correct spatial scaling even outside this limit.
Besides, $D_A$ also affects $S_B^{\mathrm{white}}(q)$ through the intensity of the white reference noise $\Xi^{\mathrm{white}}$. This intensity is proportional to the zero-frequency fluctuation spectrum of $A$, which combines the equal-time magnitude of its density fluctuations with their temporal persistence (Eq.~\eqref{eq:theory_colored_noise}). At fixed $S_A(q)$, increasing $D_A$ shortens this persistence without changing the fluctuation magnitude. The local activity regulation of $B$ then changes more rapidly, reducing its persistent effect on density fluctuations in $B$. The shorter driver persistence therefore lowers the white reference intensity and hence $S_B^{\mathrm{white}}(q)$. Note that increasing $D_A$ also raises the temporal correction factor towards unity, while the full regulation-induced contribution still decreases when both factors are included, at fixed coefficients of the $B$ dynamics.

At fixed diffusivities and coupling parameters, larger equal-time spatial correlation of drivers, $S_A(q)$, strengthens activity regulation, increasing the low-$q$ intensity of $\Xi^{\mathrm{white}}$ and the plateau in $S_B(q)$. Conversely, when $S_A(q \to 0)\sim q^2$, the induced contribution has the same scaling and $B$ remains hyperuniform. Thus, the driver's static structure determines the spatial dependence of the regulation, while its dynamics determine how persistently it acts on $B$.
Moreover, We emphasize that the effective noise $\Xi$ produces such nonhyperuniform $S(q\to 0)$ plateau without violating COM conservation, since its Laplacian form $\nabla^2\delta\rho_A$ ensures that the associated current has a vanishing spatial integral in every realization (see SI). Thus, spatiotemporally correlated effective noise can destroy hyperuniformity while preserving the COM of $B$, demonstrating that COM conservation alone does not guarantee hyperuniformity.


In conclusion, using particle simulations and a linear fluctuating theory, we have shown that COM conservation alone does not guarantee hyperuniformity in activity-regulated BRO. Spatiotemporally correlated activity can induce long-wavelength density fluctuations despite reciprocal particle displacements. This finding clarifies the limits of COM conservation as a route to hyperuniformity and shows that the large-scale structure also depends on the correlations of the driving activity. Our results suggest that designing nonequilibrium hyperuniform mixtures requires control of both particle rearrangements and the spatial and temporal correlations of activity.

\begin{acknowledgments}
This work was financially supported by the Academic Research Fund from the Singapore Ministry of Education (RG151/23, and RG88/25) and the National Research Foundation, Singapore, under its 29th Competitive Research Program (CRP) Call (NRF-CRP29-2022-0002).
\end{acknowledgments}

\paragraph{Data availability}
The data that support the findings of
this article are not publicly available. The data are available
from the authors upon reasonable request.

\bibliography{references}

\begin{thebibliography}{45}%
\makeatletter
\providecommand \@ifxundefined [1]{%
 \@ifx{#1\undefined}
}%
\providecommand \@ifnum [1]{%
 \ifnum #1\expandafter \@firstoftwo
 \else \expandafter \@secondoftwo
 \fi
}%
\providecommand \@ifx [1]{%
 \ifx #1\expandafter \@firstoftwo
 \else \expandafter \@secondoftwo
 \fi
}%
\providecommand \natexlab [1]{#1}%
\providecommand \enquote  [1]{``#1''}%
\providecommand \bibnamefont  [1]{#1}%
\providecommand \bibfnamefont [1]{#1}%
\providecommand \citenamefont [1]{#1}%
\providecommand \href@noop [0]{\@secondoftwo}%
\providecommand \href [0]{\begingroup \@sanitize@url \@href}%
\providecommand \@href[1]{\@@startlink{#1}\@@href}%
\providecommand \@@href[1]{\endgroup#1\@@endlink}%
\providecommand \@sanitize@url [0]{\catcode `\\12\catcode `\$12\catcode
  `\&12\catcode `\#12\catcode `\^12\catcode `\_12\catcode `\%12\relax}%
\providecommand \@@startlink[1]{}%
\providecommand \@@endlink[0]{}%
\providecommand \url  [0]{\begingroup\@sanitize@url \@url }%
\providecommand \@url [1]{\endgroup\@href {#1}{\urlprefix }}%
\providecommand \urlprefix  [0]{URL }%
\providecommand \Eprint [0]{\href }%
\providecommand \doibase [0]{https://doi.org/}%
\providecommand \selectlanguage [0]{\@gobble}%
\providecommand \bibinfo  [0]{\@secondoftwo}%
\providecommand \bibfield  [0]{\@secondoftwo}%
\providecommand \translation [1]{[#1]}%
\providecommand \BibitemOpen [0]{}%
\providecommand \bibitemStop [0]{}%
\providecommand \bibitemNoStop [0]{.\EOS\space}%
\providecommand \EOS [0]{\spacefactor3000\relax}%
\providecommand \BibitemShut  [1]{\csname bibitem#1\endcsname}%
\let\auto@bib@innerbib\@empty
\bibitem [{\citenamefont {Torquato}\ and\ \citenamefont
  {Stillinger}(2003)}]{torquato2003local}%
  \BibitemOpen
  \bibfield  {author} {\bibinfo {author} {\bibfnamefont {S.}~\bibnamefont
  {Torquato}}\ and\ \bibinfo {author} {\bibfnamefont {F.~H.}\ \bibnamefont
  {Stillinger}},\ }\bibfield  {title} {\bibinfo {title} {Local density
  fluctuations, hyperuniformity, and order metrics},\ }\href@noop {} {\bibfield
   {journal} {\bibinfo  {journal} {Phys. Rev. E}\ }\textbf {\bibinfo {volume}
  {68}},\ \bibinfo {pages} {041113} (\bibinfo {year} {2003})}\BibitemShut
  {NoStop}%
\bibitem [{\citenamefont {Torquato}(2018)}]{torquato2018hyperuniform}%
  \BibitemOpen
  \bibfield  {author} {\bibinfo {author} {\bibfnamefont {S.}~\bibnamefont
  {Torquato}},\ }\bibfield  {title} {\bibinfo {title} {Hyperuniform states of
  matter},\ }\href {https://doi.org/10.1016/j.physrep.2018.03.001} {\bibfield
  {journal} {\bibinfo  {journal} {Phys. Rep.}\ }\textbf {\bibinfo {volume}
  {745}},\ \bibinfo {pages} {1} (\bibinfo {year} {2018})}\BibitemShut {NoStop}%
\bibitem [{\citenamefont {Jiao}\ \emph {et~al.}(2014)\citenamefont {Jiao},
  \citenamefont {Lau}, \citenamefont {Hatzikirou}, \citenamefont
  {Meyer-Hermann}, \citenamefont {Corbo},\ and\ \citenamefont
  {Torquato}}]{jiao2014avian}%
  \BibitemOpen
  \bibfield  {author} {\bibinfo {author} {\bibfnamefont {Y.}~\bibnamefont
  {Jiao}}, \bibinfo {author} {\bibfnamefont {T.}~\bibnamefont {Lau}}, \bibinfo
  {author} {\bibfnamefont {H.}~\bibnamefont {Hatzikirou}}, \bibinfo {author}
  {\bibfnamefont {M.}~\bibnamefont {Meyer-Hermann}}, \bibinfo {author}
  {\bibfnamefont {J.~C.}\ \bibnamefont {Corbo}},\ and\ \bibinfo {author}
  {\bibfnamefont {S.}~\bibnamefont {Torquato}},\ }\bibfield  {title} {\bibinfo
  {title} {Avian photoreceptor patterns represent a disordered hyperuniform
  solution to a multiscale packing problem},\ }\href
  {https://doi.org/10.1103/PhysRevE.89.022721} {\bibfield  {journal} {\bibinfo
  {journal} {Phys. Rev. E}\ }\textbf {\bibinfo {volume} {89}},\ \bibinfo
  {pages} {022721} (\bibinfo {year} {2014})}\BibitemShut {NoStop}%
\bibitem [{\citenamefont {Kram}\ \emph {et~al.}(2010)\citenamefont {Kram},
  \citenamefont {Mantey},\ and\ \citenamefont {Corbo}}]{kram2010avian}%
  \BibitemOpen
  \bibfield  {author} {\bibinfo {author} {\bibfnamefont {Y.~A.}\ \bibnamefont
  {Kram}}, \bibinfo {author} {\bibfnamefont {S.}~\bibnamefont {Mantey}},\ and\
  \bibinfo {author} {\bibfnamefont {J.~C.}\ \bibnamefont {Corbo}},\ }\bibfield
  {title} {\bibinfo {title} {Avian cone photoreceptors tile the retina as five
  independent, self-organizing mosaics},\ }\href
  {https://doi.org/10.1371/journal.pone.0008992} {\bibfield  {journal}
  {\bibinfo  {journal} {PLoS ONE}\ }\textbf {\bibinfo {volume} {5}},\ \bibinfo
  {pages} {e8992} (\bibinfo {year} {2010})}\BibitemShut {NoStop}%
\bibitem [{\citenamefont {Yellott}(1983)}]{yellott1983spectral}%
  \BibitemOpen
  \bibfield  {author} {\bibinfo {author} {\bibfnamefont {J.~I.}\ \bibnamefont
  {Yellott}},\ }\bibfield  {title} {\bibinfo {title} {Spectral consequences of
  photoreceptor sampling in the rhesus retina},\ }\href
  {https://doi.org/10.1126/science.6867716} {\bibfield  {journal} {\bibinfo
  {journal} {Science}\ }\textbf {\bibinfo {volume} {221}},\ \bibinfo {pages}
  {382} (\bibinfo {year} {1983})}\BibitemShut {NoStop}%
\bibitem [{\citenamefont {Ginibre}(1965)}]{ginibre1965statistical}%
  \BibitemOpen
  \bibfield  {author} {\bibinfo {author} {\bibfnamefont {J.}~\bibnamefont
  {Ginibre}},\ }\bibfield  {title} {\bibinfo {title} {Statistical ensembles of
  complex, quaternion, and real matrices},\ }\href
  {https://doi.org/10.1063/1.1704292} {\bibfield  {journal} {\bibinfo
  {journal} {J. Math. Phys.}\ }\textbf {\bibinfo {volume} {6}},\ \bibinfo
  {pages} {440} (\bibinfo {year} {1965})}\BibitemShut {NoStop}%
\bibitem [{\citenamefont {Donev}\ \emph {et~al.}(2005)\citenamefont {Donev},
  \citenamefont {Stillinger},\ and\ \citenamefont
  {Torquato}}]{donev2005unexpected}%
  \BibitemOpen
  \bibfield  {author} {\bibinfo {author} {\bibfnamefont {A.}~\bibnamefont
  {Donev}}, \bibinfo {author} {\bibfnamefont {F.~H.}\ \bibnamefont
  {Stillinger}},\ and\ \bibinfo {author} {\bibfnamefont {S.}~\bibnamefont
  {Torquato}},\ }\bibfield  {title} {\bibinfo {title} {Unexpected density
  fluctuations in jammed disordered sphere packings},\ }\href
  {https://doi.org/10.1103/PhysRevLett.95.090604} {\bibfield  {journal}
  {\bibinfo  {journal} {Phys. Rev. Lett.}\ }\textbf {\bibinfo {volume} {95}},\
  \bibinfo {pages} {090604} (\bibinfo {year} {2005})}\BibitemShut {NoStop}%
\bibitem [{\citenamefont {Kuipers}\ and\ \citenamefont
  {Niederreiter}(2012)}]{kuipers2012uniform}%
  \BibitemOpen
  \bibfield  {author} {\bibinfo {author} {\bibfnamefont {L.}~\bibnamefont
  {Kuipers}}\ and\ \bibinfo {author} {\bibfnamefont {H.}~\bibnamefont
  {Niederreiter}},\ }\href@noop {} {\emph {\bibinfo {title} {Uniform
  Distribution of Sequences}}}\ (\bibinfo  {publisher} {Courier Corporation},\
  \bibinfo {year} {2012})\BibitemShut {NoStop}%
\bibitem [{\citenamefont {Gabrielli}\ \emph {et~al.}(2002)\citenamefont
  {Gabrielli}, \citenamefont {Joyce},\ and\ \citenamefont {{Sylos
  Labini}}}]{gabrielli2002glass}%
  \BibitemOpen
  \bibfield  {author} {\bibinfo {author} {\bibfnamefont {A.}~\bibnamefont
  {Gabrielli}}, \bibinfo {author} {\bibfnamefont {M.}~\bibnamefont {Joyce}},\
  and\ \bibinfo {author} {\bibfnamefont {F.}~\bibnamefont {{Sylos Labini}}},\
  }\bibfield  {title} {\bibinfo {title} {Glass-like universe: Real-space
  correlation properties of standard cosmological models},\ }\href
  {https://doi.org/10.1103/PhysRevD.65.083523} {\bibfield  {journal} {\bibinfo
  {journal} {Phys. Rev. D}\ }\textbf {\bibinfo {volume} {65}},\ \bibinfo
  {pages} {083523} (\bibinfo {year} {2002})}\BibitemShut {NoStop}%
\bibitem [{\citenamefont {Florescu}\ \emph {et~al.}(2009)\citenamefont
  {Florescu}, \citenamefont {Torquato},\ and\ \citenamefont
  {Steinhardt}}]{florescu2009designer}%
  \BibitemOpen
  \bibfield  {author} {\bibinfo {author} {\bibfnamefont {M.}~\bibnamefont
  {Florescu}}, \bibinfo {author} {\bibfnamefont {S.}~\bibnamefont {Torquato}},\
  and\ \bibinfo {author} {\bibfnamefont {P.~J.}\ \bibnamefont {Steinhardt}},\
  }\bibfield  {title} {\bibinfo {title} {Designer disordered materials with
  large, complete photonic band gaps},\ }\href@noop {} {\bibfield  {journal}
  {\bibinfo  {journal} {Proc. Natl. Acad. Sci. U.S.A.}\ }\textbf {\bibinfo
  {volume} {106}},\ \bibinfo {pages} {20658} (\bibinfo {year}
  {2009})}\BibitemShut {NoStop}%
\bibitem [{\citenamefont {Man}\ \emph {et~al.}(2013)\citenamefont {Man},
  \citenamefont {Florescu}, \citenamefont {Williamson}, \citenamefont {He},
  \citenamefont {Hashemizad}, \citenamefont {Leung}, \citenamefont {Liner},
  \citenamefont {Torquato}, \citenamefont {Chaikin},\ and\ \citenamefont
  {Steinhardt}}]{man2013isotropic}%
  \BibitemOpen
  \bibfield  {author} {\bibinfo {author} {\bibfnamefont {W.}~\bibnamefont
  {Man}}, \bibinfo {author} {\bibfnamefont {M.}~\bibnamefont {Florescu}},
  \bibinfo {author} {\bibfnamefont {E.~P.}\ \bibnamefont {Williamson}},
  \bibinfo {author} {\bibfnamefont {Y.}~\bibnamefont {He}}, \bibinfo {author}
  {\bibfnamefont {S.~R.}\ \bibnamefont {Hashemizad}}, \bibinfo {author}
  {\bibfnamefont {B.~Y.}\ \bibnamefont {Leung}}, \bibinfo {author}
  {\bibfnamefont {D.~R.}\ \bibnamefont {Liner}}, \bibinfo {author}
  {\bibfnamefont {S.}~\bibnamefont {Torquato}}, \bibinfo {author}
  {\bibfnamefont {P.~M.}\ \bibnamefont {Chaikin}},\ and\ \bibinfo {author}
  {\bibfnamefont {P.~J.}\ \bibnamefont {Steinhardt}},\ }\bibfield  {title}
  {\bibinfo {title} {Isotropic band gaps and freeform waveguides observed in
  hyperuniform disordered photonic solids},\ }\href@noop {} {\bibfield
  {journal} {\bibinfo  {journal} {Proc. Natl. Acad. Sci. U.S.A.}\ }\textbf
  {\bibinfo {volume} {110}},\ \bibinfo {pages} {15886} (\bibinfo {year}
  {2013})}\BibitemShut {NoStop}%
\bibitem [{\citenamefont {Leseur}\ \emph {et~al.}(2016)\citenamefont {Leseur},
  \citenamefont {Pierrat},\ and\ \citenamefont {Carminati}}]{leseur2016high}%
  \BibitemOpen
  \bibfield  {author} {\bibinfo {author} {\bibfnamefont {O.}~\bibnamefont
  {Leseur}}, \bibinfo {author} {\bibfnamefont {R.}~\bibnamefont {Pierrat}},\
  and\ \bibinfo {author} {\bibfnamefont {R.}~\bibnamefont {Carminati}},\
  }\bibfield  {title} {\bibinfo {title} {High-density hyperuniform materials
  can be transparent},\ }\href@noop {} {\bibfield  {journal} {\bibinfo
  {journal} {Optica}\ }\textbf {\bibinfo {volume} {3}},\ \bibinfo {pages} {763}
  (\bibinfo {year} {2016})}\BibitemShut {NoStop}%
\bibitem [{\citenamefont {Diamant}\ and\ \citenamefont
  {O{\v{g}}uz}(2025)}]{diamant2025fluids}%
  \BibitemOpen
  \bibfield  {author} {\bibinfo {author} {\bibfnamefont {H.}~\bibnamefont
  {Diamant}}\ and\ \bibinfo {author} {\bibfnamefont {E.~C.}\ \bibnamefont
  {O{\v{g}}uz}},\ }\bibfield  {title} {\bibinfo {title} {Fluids with power-law
  repulsion: Hyperuniformity and energy fluctuations},\ }\href
  {https://doi.org/10.1088/1361-648X/ad912d} {\bibfield  {journal} {\bibinfo
  {journal} {J. Phys.: Condens. Matter}\ }\textbf {\bibinfo {volume} {37}},\
  \bibinfo {pages} {065101} (\bibinfo {year} {2025})}\BibitemShut {NoStop}%
\bibitem [{\citenamefont {Wang}\ and\ \citenamefont
  {Torquato}(2024)}]{wang2024designer}%
  \BibitemOpen
  \bibfield  {author} {\bibinfo {author} {\bibfnamefont {H.}~\bibnamefont
  {Wang}}\ and\ \bibinfo {author} {\bibfnamefont {S.}~\bibnamefont
  {Torquato}},\ }\bibfield  {title} {\bibinfo {title} {Designer pair statistics
  of disordered many-particle systems with novel properties},\ }\href
  {https://doi.org/10.1063/5.0189769} {\bibfield  {journal} {\bibinfo
  {journal} {J. Chem. Phys.}\ }\textbf {\bibinfo {volume} {160}},\ \bibinfo
  {pages} {044911} (\bibinfo {year} {2024})}\BibitemShut {NoStop}%
\bibitem [{\citenamefont {Zhang}\ and\ \citenamefont
  {Torquato}(2020)}]{Zhang2020Realizable}%
  \BibitemOpen
  \bibfield  {author} {\bibinfo {author} {\bibfnamefont {G.}~\bibnamefont
  {Zhang}}\ and\ \bibinfo {author} {\bibfnamefont {S.}~\bibnamefont
  {Torquato}},\ }\bibfield  {title} {\bibinfo {title} {Realizable hyperuniform
  and nonhyperuniform particle configurations with targeted spectral functions
  via effective pair interactions},\ }\href
  {https://doi.org/10.1103/PhysRevE.101.032124} {\bibfield  {journal} {\bibinfo
   {journal} {Physical Review E}\ }\textbf {\bibinfo {volume} {101}},\ \bibinfo
  {pages} {032124} (\bibinfo {year} {2020})}\BibitemShut {NoStop}%
\bibitem [{\citenamefont {Lei}\ and\ \citenamefont {Ni}(2025)}]{lei2025non}%
  \BibitemOpen
  \bibfield  {author} {\bibinfo {author} {\bibfnamefont {Y.}~\bibnamefont
  {Lei}}\ and\ \bibinfo {author} {\bibfnamefont {R.}~\bibnamefont {Ni}},\
  }\bibfield  {title} {\bibinfo {title} {Non-equilibrium dynamic hyperuniform
  states},\ }\href {https://doi.org/10.1088/1361-648X/ad83a0} {\bibfield
  {journal} {\bibinfo  {journal} {J. Phys.: Condens. Matter}\ }\textbf
  {\bibinfo {volume} {37}},\ \bibinfo {pages} {023004} (\bibinfo {year}
  {2025})}\BibitemShut {NoStop}%
\bibitem [{\citenamefont {Pine}\ \emph {et~al.}(2005)\citenamefont {Pine},
  \citenamefont {Gollub}, \citenamefont {Brady},\ and\ \citenamefont
  {Leshansky}}]{pine2005chaos}%
  \BibitemOpen
  \bibfield  {author} {\bibinfo {author} {\bibfnamefont {D.~J.}\ \bibnamefont
  {Pine}}, \bibinfo {author} {\bibfnamefont {J.~P.}\ \bibnamefont {Gollub}},
  \bibinfo {author} {\bibfnamefont {J.~F.}\ \bibnamefont {Brady}},\ and\
  \bibinfo {author} {\bibfnamefont {A.~M.}\ \bibnamefont {Leshansky}},\
  }\bibfield  {title} {\bibinfo {title} {Chaos and threshold for
  irreversibility in sheared suspensions},\ }\href
  {https://doi.org/10.1038/nature04380} {\bibfield  {journal} {\bibinfo
  {journal} {Nature (London)}\ }\textbf {\bibinfo {volume} {438}},\ \bibinfo
  {pages} {997} (\bibinfo {year} {2005})}\BibitemShut {NoStop}%
\bibitem [{\citenamefont {Cort{\'e}}\ \emph {et~al.}(2008)\citenamefont
  {Cort{\'e}}, \citenamefont {Chaikin}, \citenamefont {Gollub},\ and\
  \citenamefont {Pine}}]{corte2008random}%
  \BibitemOpen
  \bibfield  {author} {\bibinfo {author} {\bibfnamefont {L.}~\bibnamefont
  {Cort{\'e}}}, \bibinfo {author} {\bibfnamefont {P.~M.}\ \bibnamefont
  {Chaikin}}, \bibinfo {author} {\bibfnamefont {J.~P.}\ \bibnamefont
  {Gollub}},\ and\ \bibinfo {author} {\bibfnamefont {D.~J.}\ \bibnamefont
  {Pine}},\ }\bibfield  {title} {\bibinfo {title} {Random organization in
  periodically driven systems},\ }\href {https://doi.org/10.1038/nphys891}
  {\bibfield  {journal} {\bibinfo  {journal} {Nat. Phys.}\ }\textbf {\bibinfo
  {volume} {4}},\ \bibinfo {pages} {420} (\bibinfo {year} {2008})}\BibitemShut
  {NoStop}%
\bibitem [{\citenamefont {Hexner}\ and\ \citenamefont
  {Levine}(2015)}]{hexner2015hyperuniformity}%
  \BibitemOpen
  \bibfield  {author} {\bibinfo {author} {\bibfnamefont {D.}~\bibnamefont
  {Hexner}}\ and\ \bibinfo {author} {\bibfnamefont {D.}~\bibnamefont
  {Levine}},\ }\bibfield  {title} {\bibinfo {title} {Hyperuniformity of
  critical absorbing states},\ }\href@noop {} {\bibfield  {journal} {\bibinfo
  {journal} {Phys. Rev. Lett.}\ }\textbf {\bibinfo {volume} {114}},\ \bibinfo
  {pages} {110602} (\bibinfo {year} {2015})}\BibitemShut {NoStop}%
\bibitem [{\citenamefont {Tjhung}\ and\ \citenamefont
  {Berthier}(2015)}]{tjhung2015hyperuniform}%
  \BibitemOpen
  \bibfield  {author} {\bibinfo {author} {\bibfnamefont {E.}~\bibnamefont
  {Tjhung}}\ and\ \bibinfo {author} {\bibfnamefont {L.}~\bibnamefont
  {Berthier}},\ }\bibfield  {title} {\bibinfo {title} {Hyperuniform density
  fluctuations and diverging dynamic correlations in periodically driven
  colloidal suspensions},\ }\href@noop {} {\bibfield  {journal} {\bibinfo
  {journal} {Phys. Rev. Lett.}\ }\textbf {\bibinfo {volume} {114}},\ \bibinfo
  {pages} {148301} (\bibinfo {year} {2015})}\BibitemShut {NoStop}%
\bibitem [{\citenamefont {Ma}\ and\ \citenamefont
  {Torquato}(2019)}]{ma2019generalized}%
  \BibitemOpen
  \bibfield  {author} {\bibinfo {author} {\bibfnamefont {Z.}~\bibnamefont
  {Ma}}\ and\ \bibinfo {author} {\bibfnamefont {S.}~\bibnamefont {Torquato}},\
  }\bibfield  {title} {\bibinfo {title} {Hyperuniformity of generalized random
  organization models},\ }\href {https://doi.org/10.1103/PhysRevE.99.022115}
  {\bibfield  {journal} {\bibinfo  {journal} {Phys. Rev. E}\ }\textbf {\bibinfo
  {volume} {99}},\ \bibinfo {pages} {022115} (\bibinfo {year}
  {2019})}\BibitemShut {NoStop}%
\bibitem [{\citenamefont {Wilken}\ \emph {et~al.}(2020)\citenamefont {Wilken},
  \citenamefont {Guerra}, \citenamefont {Pine},\ and\ \citenamefont
  {Chaikin}}]{wilken2020hyperuniform}%
  \BibitemOpen
  \bibfield  {author} {\bibinfo {author} {\bibfnamefont {S.}~\bibnamefont
  {Wilken}}, \bibinfo {author} {\bibfnamefont {R.~E.}\ \bibnamefont {Guerra}},
  \bibinfo {author} {\bibfnamefont {D.~J.}\ \bibnamefont {Pine}},\ and\
  \bibinfo {author} {\bibfnamefont {P.~M.}\ \bibnamefont {Chaikin}},\
  }\bibfield  {title} {\bibinfo {title} {Hyperuniform structures formed by
  shearing colloidal suspensions},\ }\href
  {https://doi.org/10.1103/PhysRevLett.125.148001} {\bibfield  {journal}
  {\bibinfo  {journal} {Phys. Rev. Lett.}\ }\textbf {\bibinfo {volume} {125}},\
  \bibinfo {pages} {148001} (\bibinfo {year} {2020})}\BibitemShut {NoStop}%
\bibitem [{\citenamefont {Chen}\ \emph {et~al.}(2024)\citenamefont {Chen},
  \citenamefont {Lei}, \citenamefont {Xiang}, \citenamefont {Duan},
  \citenamefont {Peng},\ and\ \citenamefont {Zhang}}]{chen2024emergent}%
  \BibitemOpen
  \bibfield  {author} {\bibinfo {author} {\bibfnamefont {J.}~\bibnamefont
  {Chen}}, \bibinfo {author} {\bibfnamefont {X.}~\bibnamefont {Lei}}, \bibinfo
  {author} {\bibfnamefont {Y.}~\bibnamefont {Xiang}}, \bibinfo {author}
  {\bibfnamefont {M.}~\bibnamefont {Duan}}, \bibinfo {author} {\bibfnamefont
  {X.}~\bibnamefont {Peng}},\ and\ \bibinfo {author} {\bibfnamefont {H.~P.}\
  \bibnamefont {Zhang}},\ }\bibfield  {title} {\bibinfo {title} {Emergent
  chirality and hyperuniformity in an active mixture with nonreciprocal
  interactions},\ }\href {https://doi.org/10.1103/PhysRevLett.132.118301}
  {\bibfield  {journal} {\bibinfo  {journal} {Phys. Rev. Lett.}\ }\textbf
  {\bibinfo {volume} {132}},\ \bibinfo {pages} {118301} (\bibinfo {year}
  {2024})}\BibitemShut {NoStop}%
\bibitem [{\citenamefont {Lei}\ and\ \citenamefont
  {Ni}(2023)}]{lei2023howHUfreeze}%
  \BibitemOpen
  \bibfield  {author} {\bibinfo {author} {\bibfnamefont {Y.}~\bibnamefont
  {Lei}}\ and\ \bibinfo {author} {\bibfnamefont {R.}~\bibnamefont {Ni}},\
  }\bibfield  {title} {\bibinfo {title} {How does a hyperuniform fluid
  freeze?},\ }\href@noop {} {\bibfield  {journal} {\bibinfo  {journal} {Proc.
  Natl. Acad. Sci. U.S.A.}\ }\textbf {\bibinfo {volume} {120}},\ \bibinfo
  {pages} {e2312866120} (\bibinfo {year} {2023})}\BibitemShut {NoStop}%
\bibitem [{\citenamefont {Lei}\ and\ \citenamefont
  {Ni}(2026)}]{lei2026metastable}%
  \BibitemOpen
  \bibfield  {author} {\bibinfo {author} {\bibfnamefont {Y.}~\bibnamefont
  {Lei}}\ and\ \bibinfo {author} {\bibfnamefont {R.}~\bibnamefont {Ni}},\
  }\href@noop {} {\bibinfo {title} {Metastable hyperuniformity at discontinuous
  absorbing transitions}} (\bibinfo {year} {2026}),\ \Eprint
  {https://arxiv.org/abs/2605.13561} {arXiv:2605.13561 [cond-mat.soft]}
  \BibitemShut {NoStop}%
\bibitem [{\citenamefont {Hexner}\ and\ \citenamefont
  {Levine}(2017)}]{hexner2017noise}%
  \BibitemOpen
  \bibfield  {author} {\bibinfo {author} {\bibfnamefont {D.}~\bibnamefont
  {Hexner}}\ and\ \bibinfo {author} {\bibfnamefont {D.}~\bibnamefont
  {Levine}},\ }\bibfield  {title} {\bibinfo {title} {Noise, diffusion, and
  hyperuniformity},\ }\href@noop {} {\bibfield  {journal} {\bibinfo  {journal}
  {Phys. Rev. Lett.}\ }\textbf {\bibinfo {volume} {118}},\ \bibinfo {pages}
  {020601} (\bibinfo {year} {2017})}\BibitemShut {NoStop}%
\bibitem [{\citenamefont {Anand}\ \emph {et~al.}(2026)\citenamefont {Anand},
  \citenamefont {Zhang},\ and\ \citenamefont
  {Martiniani}}]{anand2026universal}%
  \BibitemOpen
  \bibfield  {author} {\bibinfo {author} {\bibfnamefont {S.}~\bibnamefont
  {Anand}}, \bibinfo {author} {\bibfnamefont {G.}~\bibnamefont {Zhang}},\ and\
  \bibinfo {author} {\bibfnamefont {S.}~\bibnamefont {Martiniani}},\ }\bibfield
   {title} {\bibinfo {title} {Emergent universal long-range structure in
  random-organizing systems},\ }\href
  {https://doi.org/10.1038/s41467-026-68601-2} {\bibfield  {journal} {\bibinfo
  {journal} {Nat. Commun.}\ }\textbf {\bibinfo {volume} {17}},\ \bibinfo
  {pages} {2346} (\bibinfo {year} {2026})}\BibitemShut {NoStop}%
\bibitem [{\citenamefont {Lei}\ and\ \citenamefont
  {Ni}(2019)}]{lei2019hydrodynamics}%
  \BibitemOpen
  \bibfield  {author} {\bibinfo {author} {\bibfnamefont {Q.-L.}\ \bibnamefont
  {Lei}}\ and\ \bibinfo {author} {\bibfnamefont {R.}~\bibnamefont {Ni}},\
  }\bibfield  {title} {\bibinfo {title} {Hydrodynamics of random-organizing
  hyperuniform fluids},\ }\href@noop {} {\bibfield  {journal} {\bibinfo
  {journal} {Proc. Natl. Acad. Sci. U.S.A.}\ }\textbf {\bibinfo {volume}
  {116}},\ \bibinfo {pages} {22983} (\bibinfo {year} {2019})}\BibitemShut
  {NoStop}%
\bibitem [{\citenamefont {Li}\ \emph {et~al.}(2025)\citenamefont {Li},
  \citenamefont {Lei},\ and\ \citenamefont {Ma}}]{li2025fluidization}%
  \BibitemOpen
  \bibfield  {author} {\bibinfo {author} {\bibfnamefont {Z.-Q.}\ \bibnamefont
  {Li}}, \bibinfo {author} {\bibfnamefont {Q.-L.}\ \bibnamefont {Lei}},\ and\
  \bibinfo {author} {\bibfnamefont {Y.-Q.}\ \bibnamefont {Ma}},\ }\bibfield
  {title} {\bibinfo {title} {Fluidization and anomalous density fluctuations in
  {2D} {Voronoi} cell tissues with pulsating activity},\ }\href
  {https://doi.org/10.1073/pnas.2421518122} {\bibfield  {journal} {\bibinfo
  {journal} {Proc. Natl. Acad. Sci. U.S.A.}\ }\textbf {\bibinfo {volume}
  {122}},\ \bibinfo {pages} {e2421518122} (\bibinfo {year} {2025})}\BibitemShut
  {NoStop}%
\bibitem [{\citenamefont {Lei}\ \emph {et~al.}(2019)\citenamefont {Lei},
  \citenamefont {{Pica Ciamarra}},\ and\ \citenamefont
  {Ni}}]{lei2019nonequilibrium}%
  \BibitemOpen
  \bibfield  {author} {\bibinfo {author} {\bibfnamefont {Q.-L.}\ \bibnamefont
  {Lei}}, \bibinfo {author} {\bibfnamefont {M.}~\bibnamefont {{Pica
  Ciamarra}}},\ and\ \bibinfo {author} {\bibfnamefont {R.}~\bibnamefont {Ni}},\
  }\bibfield  {title} {\bibinfo {title} {Nonequilibrium strongly hyperuniform
  fluids of circle active particles with large local density fluctuations},\
  }\href@noop {} {\bibfield  {journal} {\bibinfo  {journal} {Sci. Adv.}\
  }\textbf {\bibinfo {volume} {5}},\ \bibinfo {pages} {eaau7423} (\bibinfo
  {year} {2019})}\BibitemShut {NoStop}%
\bibitem [{\citenamefont {Kuroda}\ and\ \citenamefont
  {Miyazaki}(2023)}]{kuroda2023microscopic}%
  \BibitemOpen
  \bibfield  {author} {\bibinfo {author} {\bibfnamefont {Y.}~\bibnamefont
  {Kuroda}}\ and\ \bibinfo {author} {\bibfnamefont {K.}~\bibnamefont
  {Miyazaki}},\ }\bibfield  {title} {\bibinfo {title} {Microscopic theory for
  hyperuniformity in two-dimensional chiral active fluid},\ }\href
  {https://doi.org/10.1088/1742-5468/ad0639} {\bibfield  {journal} {\bibinfo
  {journal} {J. Stat. Mech.: Theory Exp.}\ }\textbf {\bibinfo {volume}
  {2023}}\bibinfo  {number} { (10)},\ \bibinfo {pages} {103203}}\BibitemShut
  {NoStop}%
\bibitem [{\citenamefont {Zhang}\ and\ \citenamefont
  {Snezhko}(2022)}]{zhangHyperuniformActiveChiral2022}%
  \BibitemOpen
\bibfield  {number} {  }\bibfield  {author} {\bibinfo {author} {\bibfnamefont
  {B.}~\bibnamefont {Zhang}}\ and\ \bibinfo {author} {\bibfnamefont
  {A.}~\bibnamefont {Snezhko}},\ }\bibfield  {title} {\bibinfo {title}
  {Hyperuniform active chiral fluids with tunable internal structure},\ }\href
  {https://doi.org/10.1103/PhysRevLett.128.218002} {\bibfield  {journal}
  {\bibinfo  {journal} {Phys. Rev. Lett.}\ }\textbf {\bibinfo {volume} {128}},\
  \bibinfo {pages} {218002} (\bibinfo {year} {2022})}\BibitemShut {NoStop}%
\bibitem [{\citenamefont {De~Luca}\ \emph {et~al.}(2024)\citenamefont
  {De~Luca}, \citenamefont {Ma}, \citenamefont {Nardini},\ and\ \citenamefont
  {Cates}}]{delucaHyperuniformityPhaseOrdering2024a}%
  \BibitemOpen
  \bibfield  {author} {\bibinfo {author} {\bibfnamefont {F.}~\bibnamefont
  {De~Luca}}, \bibinfo {author} {\bibfnamefont {X.}~\bibnamefont {Ma}},
  \bibinfo {author} {\bibfnamefont {C.}~\bibnamefont {Nardini}},\ and\ \bibinfo
  {author} {\bibfnamefont {M.~E.}\ \bibnamefont {Cates}},\ }\bibfield  {title}
  {\bibinfo {title} {Hyperuniformity in phase ordering: the roles of activity,
  noise, and non-constant mobility},\ }\href
  {https://doi.org/10.1088/1361-648X/ad5b45} {\bibfield  {journal} {\bibinfo
  {journal} {J. Phys.: Condens. Matter}\ }\textbf {\bibinfo {volume} {36}},\
  \bibinfo {pages} {405101} (\bibinfo {year} {2024})}\BibitemShut {NoStop}%
\bibitem [{\citenamefont {Maire}\ and\ \citenamefont
  {Chaix}(2025)}]{maire2025hyperuniformity}%
  \BibitemOpen
  \bibfield  {author} {\bibinfo {author} {\bibfnamefont {R.}~\bibnamefont
  {Maire}}\ and\ \bibinfo {author} {\bibfnamefont {L.}~\bibnamefont {Chaix}},\
  }\bibfield  {title} {\bibinfo {title} {Hyperuniformity and conservation laws
  in non-equilibrium systems},\ }\href {https://doi.org/10.1063/5.0300881}
  {\bibfield  {journal} {\bibinfo  {journal} {J. Chem. Phys.}\ }\textbf
  {\bibinfo {volume} {163}},\ \bibinfo {pages} {214507} (\bibinfo {year}
  {2025})}\BibitemShut {NoStop}%
\bibitem [{\citenamefont {Casiulis}\ \emph {et~al.}(2026)\citenamefont
  {Casiulis}, \citenamefont {Anand},\ and\ \citenamefont
  {Martiniani}}]{casiulis2026hyperuniform}%
  \BibitemOpen
  \bibfield  {author} {\bibinfo {author} {\bibfnamefont {M.}~\bibnamefont
  {Casiulis}}, \bibinfo {author} {\bibfnamefont {S.}~\bibnamefont {Anand}},\
  and\ \bibinfo {author} {\bibfnamefont {S.}~\bibnamefont {Martiniani}},\
  }\href {https://doi.org/10.48550/arXiv.2607.07411} {\bibinfo {title}
  {Hyperuniform systems are maximally irreversible}} (\bibinfo {year} {2026}),\
  \Eprint {https://arxiv.org/abs/2607.07411} {arXiv:2607.07411 [cond-mat.soft]}
  \BibitemShut {NoStop}%
\bibitem [{\citenamefont {Lei}\ \emph {et~al.}(2023)\citenamefont {Lei},
  \citenamefont {Zheng},\ and\ \citenamefont {Ni}}]{lei2023sphere}%
  \BibitemOpen
  \bibfield  {author} {\bibinfo {author} {\bibfnamefont {Y.}~\bibnamefont
  {Lei}}, \bibinfo {author} {\bibfnamefont {N.}~\bibnamefont {Zheng}},\ and\
  \bibinfo {author} {\bibfnamefont {R.}~\bibnamefont {Ni}},\ }\bibfield
  {title} {\bibinfo {title} {Random organization and non-equilibrium
  hyperuniform fluids on a sphere},\ }\href {https://doi.org/10.1063/5.0165527}
  {\bibfield  {journal} {\bibinfo  {journal} {J. Chem. Phys.}\ }\textbf
  {\bibinfo {volume} {159}},\ \bibinfo {pages} {081101} (\bibinfo {year}
  {2023})}\BibitemShut {NoStop}%
\bibitem [{\citenamefont {Maire}\ and\ \citenamefont
  {Plati}(2024)}]{maire2024enhancing}%
  \BibitemOpen
  \bibfield  {author} {\bibinfo {author} {\bibfnamefont {R.}~\bibnamefont
  {Maire}}\ and\ \bibinfo {author} {\bibfnamefont {A.}~\bibnamefont {Plati}},\
  }\bibfield  {title} {\bibinfo {title} {Enhancing (quasi-)long-range order in
  a two-dimensional driven crystal},\ }\href
  {https://doi.org/10.1063/5.0217958} {\bibfield  {journal} {\bibinfo
  {journal} {J. Chem. Phys.}\ }\textbf {\bibinfo {volume} {161}},\ \bibinfo
  {pages} {054902} (\bibinfo {year} {2024})}\BibitemShut {NoStop}%
\bibitem [{\citenamefont {Galliano}\ \emph {et~al.}(2023)\citenamefont
  {Galliano}, \citenamefont {Cates},\ and\ \citenamefont
  {Berthier}}]{galliano2023crystals}%
  \BibitemOpen
  \bibfield  {author} {\bibinfo {author} {\bibfnamefont {L.}~\bibnamefont
  {Galliano}}, \bibinfo {author} {\bibfnamefont {M.~E.}\ \bibnamefont
  {Cates}},\ and\ \bibinfo {author} {\bibfnamefont {L.}~\bibnamefont
  {Berthier}},\ }\bibfield  {title} {\bibinfo {title} {Two-dimensional crystals
  far from equilibrium},\ }\href
  {https://doi.org/10.1103/PhysRevLett.131.047101} {\bibfield  {journal}
  {\bibinfo  {journal} {Phys. Rev. Lett.}\ }\textbf {\bibinfo {volume} {131}},\
  \bibinfo {pages} {047101} (\bibinfo {year} {2023})}\BibitemShut {NoStop}%
\bibitem [{\citenamefont {Maire}\ \emph
  {et~al.}(2025{\natexlab{a}})\citenamefont {Maire}, \citenamefont {Galliano},
  \citenamefont {Plati},\ and\ \citenamefont {Berthier}}]{maire2025interfaces}%
  \BibitemOpen
  \bibfield  {author} {\bibinfo {author} {\bibfnamefont {R.}~\bibnamefont
  {Maire}}, \bibinfo {author} {\bibfnamefont {L.}~\bibnamefont {Galliano}},
  \bibinfo {author} {\bibfnamefont {A.}~\bibnamefont {Plati}},\ and\ \bibinfo
  {author} {\bibfnamefont {L.}~\bibnamefont {Berthier}},\ }\bibfield  {title}
  {\bibinfo {title} {Hyperuniform interfaces in nonequilibrium phase
  coexistence},\ }\href {https://doi.org/10.1103/4b8v-4sbh} {\bibfield
  {journal} {\bibinfo  {journal} {Phys. Rev. Lett.}\ }\textbf {\bibinfo
  {volume} {135}},\ \bibinfo {pages} {227102} (\bibinfo {year}
  {2025}{\natexlab{a}})}\BibitemShut {NoStop}%
\bibitem [{\citenamefont {Maire}\ \emph
  {et~al.}(2025{\natexlab{b}})\citenamefont {Maire}, \citenamefont {Plati},
  \citenamefont {Smallenburg},\ and\ \citenamefont
  {Foffi}}]{maire2025granular}%
  \BibitemOpen
  \bibfield  {author} {\bibinfo {author} {\bibfnamefont {R.}~\bibnamefont
  {Maire}}, \bibinfo {author} {\bibfnamefont {A.}~\bibnamefont {Plati}},
  \bibinfo {author} {\bibfnamefont {F.}~\bibnamefont {Smallenburg}},\ and\
  \bibinfo {author} {\bibfnamefont {G.}~\bibnamefont {Foffi}},\ }\bibfield
  {title} {\bibinfo {title} {Dynamical and structural properties of an
  absorbing phase transition: a case study from granular systems},\ }\href
  {https://doi.org/10.1088/1742-5468/ae23bc} {\bibfield  {journal} {\bibinfo
  {journal} {J. Stat. Mech.: Theory Exp.}\ }\textbf {\bibinfo {volume}
  {2025}}\bibinfo  {number} { (12)},\ \bibinfo {pages} {123206}}\BibitemShut
  {NoStop}%
\bibitem [{\citenamefont {Kuroda}\ \emph {et~al.}(2025)\citenamefont {Kuroda},
  \citenamefont {Kawasaki},\ and\ \citenamefont
  {Miyazaki}}]{kuroda2025crystal}%
  \BibitemOpen
\bibfield  {number} {  }\bibfield  {author} {\bibinfo {author} {\bibfnamefont
  {Y.}~\bibnamefont {Kuroda}}, \bibinfo {author} {\bibfnamefont
  {T.}~\bibnamefont {Kawasaki}},\ and\ \bibinfo {author} {\bibfnamefont
  {K.}~\bibnamefont {Miyazaki}},\ }\bibfield  {title} {\bibinfo {title}
  {Long-range translational order and hyperuniformity in two-dimensional chiral
  active crystal},\ }\href {https://doi.org/10.1103/PhysRevResearch.7.L012048}
  {\bibfield  {journal} {\bibinfo  {journal} {Phys. Rev. Research}\ }\textbf
  {\bibinfo {volume} {7}},\ \bibinfo {pages} {L012048} (\bibinfo {year}
  {2025})}\BibitemShut {NoStop}%
\bibitem [{\citenamefont {Gao}\ \emph {et~al.}(2026)\citenamefont {Gao},
  \citenamefont {Shang}, \citenamefont {Hu}, \citenamefont {Ma},\ and\
  \citenamefont {Lei}}]{gao2026criticality}%
  \BibitemOpen
  \bibfield  {author} {\bibinfo {author} {\bibfnamefont {S.}~\bibnamefont
  {Gao}}, \bibinfo {author} {\bibfnamefont {H.}~\bibnamefont {Shang}}, \bibinfo
  {author} {\bibfnamefont {H.}~\bibnamefont {Hu}}, \bibinfo {author}
  {\bibfnamefont {Y.-Q.}\ \bibnamefont {Ma}},\ and\ \bibinfo {author}
  {\bibfnamefont {Q.-L.}\ \bibnamefont {Lei}},\ }\bibfield  {title} {\bibinfo
  {title} {Liquid-gas criticality of hyperuniform fluids},\ }\href
  {https://doi.org/10.1103/3151-4dnh} {\bibfield  {journal} {\bibinfo
  {journal} {Phys. Rev. X}\ }\textbf {\bibinfo {volume} {16}},\ \bibinfo
  {pages} {021032} (\bibinfo {year} {2026})}\BibitemShut {NoStop}%
\bibitem [{\citenamefont {Maire}(2026)}]{maire2026nucleation}%
  \BibitemOpen
  \bibfield  {author} {\bibinfo {author} {\bibfnamefont {R.}~\bibnamefont
  {Maire}},\ }\bibfield  {title} {\bibinfo {title} {Hyperuniformity in active
  fluids reshapes nucleation and capillary-wave dynamics},\ }\href
  {https://doi.org/10.1103/kl3b-ct4p} {\bibfield  {journal} {\bibinfo
  {journal} {Phys. Rev. E}\ }\textbf {\bibinfo {volume} {113}},\ \bibinfo
  {pages} {064103} (\bibinfo {year} {2026})}\BibitemShut {NoStop}%
\bibitem [{\citenamefont {Bertrand}\ \emph {et~al.}(2019)\citenamefont
  {Bertrand}, \citenamefont {Chatenay},\ and\ \citenamefont
  {Voituriez}}]{bertrand2019nonlinear}%
  \BibitemOpen
  \bibfield  {author} {\bibinfo {author} {\bibfnamefont {T.}~\bibnamefont
  {Bertrand}}, \bibinfo {author} {\bibfnamefont {D.}~\bibnamefont {Chatenay}},\
  and\ \bibinfo {author} {\bibfnamefont {R.}~\bibnamefont {Voituriez}},\
  }\bibfield  {title} {\bibinfo {title} {Nonlinear diffusion and
  hyperuniformity from {Poisson} representation in systems with interaction
  mediated dynamics},\ }\href {https://doi.org/10.1088/1367-2630/ab5f17}
  {\bibfield  {journal} {\bibinfo  {journal} {New J. Phys.}\ }\textbf {\bibinfo
  {volume} {21}},\ \bibinfo {pages} {123048} (\bibinfo {year}
  {2019})}\BibitemShut {NoStop}%
\bibitem [{\citenamefont {Tjhung}\ and\ \citenamefont
  {Berthier}(2017)}]{tjhung2017discontinuous}%
  \BibitemOpen
  \bibfield  {author} {\bibinfo {author} {\bibfnamefont {E.}~\bibnamefont
  {Tjhung}}\ and\ \bibinfo {author} {\bibfnamefont {L.}~\bibnamefont
  {Berthier}},\ }\bibfield  {title} {\bibinfo {title} {Discontinuous
  fluidization transition in time-correlated assemblies of actively deforming
  particles},\ }\href {https://doi.org/10.1103/PhysRevE.96.050601} {\bibfield
  {journal} {\bibinfo  {journal} {Phys. Rev. E}\ }\textbf {\bibinfo {volume}
  {96}},\ \bibinfo {pages} {050601} (\bibinfo {year} {2017})}\BibitemShut
  {NoStop}%
\end{thebibliography}%

\end{document}